# SATELLITE-MOUNTED LIGHT SOURCES AS PHOTOMETRIC CALIBRATION STANDARDS

**Justin Albert[*], Kristie Foster, Grace Dupuis, Kyle Fransham, Kristin Koopmans[1], and Michael Jarrett**

*Dept. of Physics and Astronomy, Univ. of Victoria, Victoria, B.C., Canada*

**James Battat[2]**

*Dept. of Astronomy, Harvard Univ., Cambridge, MA, USA*

**Abstract:** A significant and growing portion of systematic error on a number of fundamental parameters in astrophysics and cosmology is due to uncertainties from absolute photometric and flux standards. A path toward achieving major reduction in such uncertainties may be provided by satellite-mounted light sources, resulting in improvement in the ability to precisely characterize atmospheric extinction, and thus helping to usher in the coming generation of precision results in astronomy. Toward this end, we have performed a campaign of observations of the 532 nm pulsed laser aboard the CALIPSO satellite, using a portable network of cameras and photodiodes, to precisely measure atmospheric extinction.

## 1. INTRODUCTION

While our understanding of the Universe has changed and improved dramatically over the past 25 years, the improvement of our knowledge of absolute spectra and flux from standard calibration sources, upon which the precision of measurements of the expansion history of the Universe [1], and of stellar and galactic evolution [2] are based, has been far slower and not kept up with reduction of other major uncertainties. As a result, uncertainties on absolute standards now constitute one of the dominant systematics for measurements such as the expansion history of the Universe using type Ia supernovae [3], and a significant systematic for measurements of stellar population in galaxy cluster counts [4], and upcoming photometric redshift surveys measuring growth of structure [5]. There are prospects for improvement in uncertainties from standard star flux and spectra [6], but the traditional techniques of measurement of standard stellar flux from above the atmosphere suffer from basic and inherent problems: the variability of all stellar sources, and the difficulty of creating a precisely calibrated, cross-checked, and stable platform for observation above the Earth's atmosphere.

The presence of an absolute flux standard in orbit above the Earth's atmosphere could provide important cross-checks and potential significant reduction of photometric and other atmospheric uncertainties for measurements that depend on such calibration. Monochromatic sources, especially ones that could cover multiple discrete wavelengths, or tune over a spectrum, could potentially help to further reduce spectrophotometric error. For calibration of telescope optics and detector characteristics, authors [7] have both conceived of and used a wavelength-tunable laser within present and upcoming telescope domes as a color calibration standard. Although a wavelength-tunable laser calibration source in orbit [8] does not exist yet, at present there is a 532 nm laser in low-Earth orbit pointed toward the Earth's surface, with precise radiometric measurement of the energy of each of the 20.25 Hz laser pulses, on the CALIPSO satellite, launched in April 2006 [9]. We have collected data from a portable network of seven cameras and two calibrated photodiodes, taken during CALIPSO flyovers on clear days in various locations in western North America. The cameras and photodiodes respectively capture images and pulses from the eye-visible green laser spot at the zenith during the moment of a flyover. Using precise pulse-by-pulse radiometry data from the CALIPSO satellite, we compare the pulse energy received on the ground with the pulse energy recorded by CALIPSO. The ratio determines the atmospheric extinction.

---

[*] Primary author, e-mail: jalbert@uvic.ca
[1] Now at the Dept. of Physics, Univ. of Regina, Regina, SK, Canada
[2] Now at the Dept. of Physics, Massachusetts Institute of Technology, Cambridge, MA, USA

Both laboratory and field calibration of the cameras and the photodiodes is essential for analysis of the images and the determinations of absolute flux. We developed a calibration system to determine the sizes of effects such as image anisotropy, nonlinearity, and temperature dependence, so that we could accurately and precisely measure optical energy of a source from its image.

## 2. SATELLITE-MOUNTED LIGHT SOURCES

Throughout history prior to 1957, the only sources of light above the Earth's atmosphere were natural in origin: stars, and reflected light from planets, moons, comets, etc. Natural sources have of course served extremely well in astronomy: through understanding the physical processes governing stellar evolution, we are now able to precisely understand the spectra of stars used as calibration sources. Nevertheless, in all stars the vast bulk of material, and the thermonuclear processes that themselves provide the light, lie beyond our sight below the surface of the star. Superb models of stellar structure are available, but uncertainties of many types always remain.

Since the launching of the first man-made satellites, a separate class of potential light sources in space has become available. Observable light from most satellites is primarily due to direct solar reflection, or reflection from Earth's albedo. While providing a convenient method of observing satellites, this light is typically unsuitable for use as a calibrated light source due to large uncertainties in the reflectivity (and, to a lesser extent, the precise orientation and reflective area) of satellites' surfaces. Reflected solar light has, however, been successfully used as an absolute infrared calibration source by the Midcourse Space Experiment (MSX), using 2 cm diameter black-coated spheres ejected from the MSX satellite, whose infrared emission was monitored by the instruments aboard MSX [10]. This technique proved highly effective for the MSX infrared calibration; however, the technique is not easily applicable to measuring extinction of visible light in the atmosphere.

Many satellites have retroreflective cubes intended for use in satellite laser ranging. Reflected laser light from retroreflectors is critical for distance measurements using precise timing; however, like solar reflection, retroreflected laser light unfortunately also suffers from uncertainties in the reflectivity of the cubes, and in reflectivity as a function of incident angle, that are too large to provide a means of measuring atmospheric extinction [11]. Thus we are left with dedicated light sources aboard satellites themselves as the sole practical means of having a satellite-based visible light source for calibration of ground-based telescopes.

Many satellites carry some means of producing observable visible light, for self-calibration purposes or otherwise. The Hubble Space Telescope is one of many satellites carrying tungsten, as well as deuterium, lamps for absolute self-calibration purposes [12]. Lamps for self-calibration are not limited to space telescopes for astronomy; earth observation and weather satellites also commonly use internal tungsten lamps as calibration sources [13]. Such internal calibration lamps are typically limited by the fact that they can degrade individually, and can be compared only with astronomical sources after launch, leaving stellar light as the only practical way to "calibrate the calibration device." Thus, such devices typically provide a cross-check rather than the basis for a true absolute irradiance calibration, or provide a means for a separate calibration, such as flat-field [14]. Furthermore, present-day internal calibration lamps aboard satellites are certainly not intended for, nor are capable of, a direct calibration of the atmospheric extinction that affects ground-based telescopes.

However, a satellite-based absolute calibration source for ground-based telescopes is not technically prohibitive. As an example, a standard household 25-watt tungsten filament lightbulb (which typically have a temperature of the order of 3000 K and usually produce approximately 1 watt of visible light between 390 and 780 nm) that radiates light equally in all directions from a 700 km low Earth orbit has an equivalent brightness to a 12.5-magnitude star (in the AB system, although for this approximate value the system makes little difference). In general, the apparent magnitude of an orbiting lamp at a typical incandescent temperature that radiates isotropically is approximately given by

$$m \approx -5.0 \log_{10}\left(\frac{\left(\ln\left(\frac{P}{1\,\text{watt}}\right)\right)^3}{h}\right) + 5.9, \tag{1}$$

where $P$ is the power of the lamp in watts, and $h$ is the height of the orbit in kilometers. The dominant uncertainties in the precise amount of light received by a ground-based telescope from such an orbiting lamp would stem from degradations of both the lamp and the power source over the lifetime of the lamp, the comparison of the spectrum of the lamp with that of typical stars, background from reflected earthshine, sunshine, moonshine, or starlight from the surface of the satellite itself, and potential deviations from perfect isotropic output of the light from the lamp.

An alternative to an isotropic or near-isotropic lamp would be a laser source, with beam pointed at the observer (with a small moveable mirror, for example). Divergences of laser beams are typically on the order of a milliradian (which can be reduced to microradians with a beam expander) so much less output power than a lamp would be required for a laser beam to mimic the brightness of a typical star. (However, note that the wall-plug power of typical diode-pumped lasers is typically in the vicinity of 20 times the output laser power [15], which is furthermore a large reduction in wall-plug power from the $10^4$ level typical of flashlamp-pumped lasers.) The apparent magnitude of an orbiting laser with Gaussian beam divergence pointed directly at a ground-based telescope is given by

$$m \approx -2.5 \log_{10}\left(\frac{P}{h^2 d^2}\right) - 20.1, \tag{2}$$

where $P$ is the laser output power in milliwatts, $h$ is the height of the orbit in kilometers, and $d$ is the RMS divergence of the laser beam in milliradians, under the assumption that the aperture of the telescope is small compared with the RMS width of the beam at the ground, $hd$. Clearly, even a sub-milliwatt laser in low Earth orbit would need to have either its divergence increased or its power reduced via filtering for it to be suitable for astronomical calibration. We shall consider the filtering option. The major uncertainties in the amount of light in the amount of light received by a ground-based telescope from an orbiting laser would stem from uncertainties in the pointing and beam profile of the laser light (which would likely need to be monitored by an array of small dedicated telescopes outboard of the main ground-based telescope), in time-dependent variation of the laser output power (which would likely need to be monitored by onboard radiometry), and in degradation of the filter over time. Nevertheless, laser light has the benefit of being monochromatic, allowing for calibration of individual wavelengths. With a widely-tunable laser, an entire spectrum could be calibrated, removing the significant inherent uncertainties associated with comparing the spectra of astrophysical objects with the spectrum of a calibration source.

The uncertainty on the apparent magnitude of an orbiting laser stemming from uncertainties in the radiometrically-monitored laser power would be limited by the precision of current radiometer technology. Modern solar radiometers, using electrical substitution radiometry, can achieve a precision of approximately 100 parts per million [16]. Uncertainties on the magnitude due to uncertainty in the pointing and beam profile would potentially be limited by the size of the array of small outboard telescopes for monitoring the laser spot, and by calibration differences between the individual telescopes in the array and with the main central telescope. The latter could clearly be minimized by a ground system for ensuring the relative calibration of the outboard telescopes and main telescope all are consistent.

The uncertainties considered above assume that the exposure time is long compared with the coherence time of the atmosphere. With short exposures, or in the case of a laser that either quickly sweeps past, or is pulsed, atmospheric scintillation can play a major role in uncertainty in apparent magnitude of a satellite-mounted source. A typical timescale for a CW laser with 1 milliradian divergence in low Earth orbit to sweep past is tens of milliseconds, which is of the same order as characteristic timescales of atmospheric scintillation, and the typical timescale of single laser pulses is nanoseconds, much shorter than scintillation timescales, thus one cannot assume that such effects can be time-averaged over. In idealized conditions, for small apertures $D < \sim 5$ cm and sub-millisecond integration times, the relative standard deviation in intensity $\sigma_I \equiv \Delta I / \langle I \rangle$, where $\Delta I$ is the root-mean-square value of $I$, is given by the square root of

$$\sigma_I^2 = 19.12 \lambda^{-7/6} \int_0^\infty C_n^2(h) h^{5/6} dh, \qquad (3)$$

where $\lambda$ is optical wavelength (in meters), $C_n^2(h)$ is known as the refractive-index structure coefficient, and $h$ is altitude (in meters) [17]. Large apertures $D > \sim 50$ cm have a relative standard deviation in intensity given by the square root of

$$\sigma_I^2 = 29.48 D^{-7/3} \int_0^\infty C_n^2(h) h^2 dh \qquad (4)$$

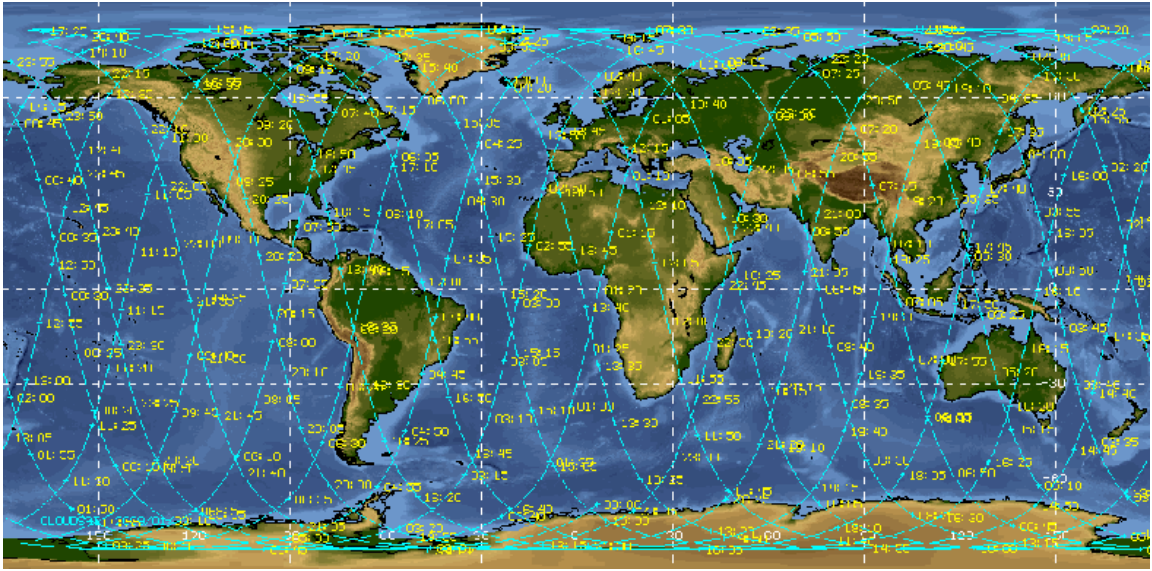

Fig. 1: A typical 1-day ground track of the CALIPSO satellite.

[17]. The values and functional form of $C_n^2(h)$ are entirely dependent on the particular atmospheric conditions at the time of observation, however a relatively typical profile is given by the Hufnagel-Valley form:

$$C_n^2(h) = 5.94 \times 10^{-53} (v/27)^2 h^{10} e^{-h/1000} + 2.7 \times 10^{-16} e^{-h/1500} + A e^{-h/100}, \qquad (5)$$

where $A$ and $v$ are free parameters [18]. Commonly-used values for the $A$ and $v$ parameters, which represent the strength of turbulence near ground level and the high-altitude wind speed respectively, are $A = 1.7 \times 10^{-14}$ m$^{-2/3}$ and $v = 21$ m/s [19]. Using these particular values, for a small aperture, the relative variance $\sigma_I^2$ would be expected to be approximately 0.22 for 532 nm light, which is not far off experimental scintillation values for a clear night at a typical location [20]. For a single small camera, this is an extremely large uncertainty. Other than by increasing integration time (which is not possible with a pulsed laser) or by significantly increasing the camera aperture, the only way to reduce this uncertainty is to increase the number of cameras. With $N$ cameras performing an observation, which are spaced further apart than the coherence length of atmospheric turbulence (typically 5 to 50 cm), the uncertainty from scintillation can be reduced by a factor $\sqrt{N}$ (for large $N$).

The analysis above considers a hypothetical pointable satellite-mounted calibration laser, and is necessarily both speculative and approximate. However, at present there is an actual laser in low Earth orbit, visible with both equipment and with the naked eye, and analysis of ground-based observational data of the laser spot can be used for comparisons with the above, as well as for development of and predictions for potential future satellite-based photometric calibration sources of ground telescopes.

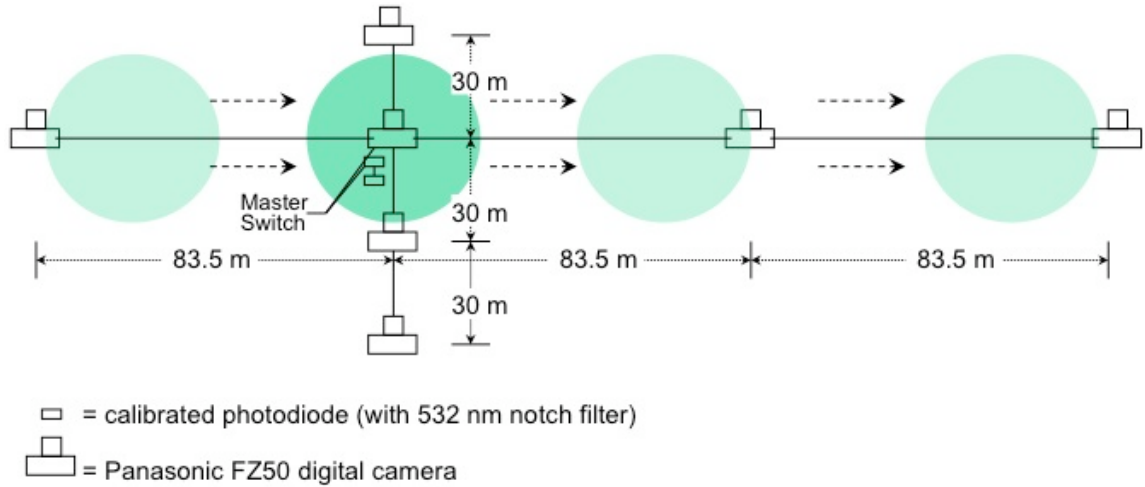

Fig. 2: CALIPSO satellite observation setup using seven Panasonic FZ50 cameras connected by coaxial cable, to obtain images of four pulses of the CALIPSO laser and obtain constraints on the width of the beam image on the ground.

## 3. THE CALIPSO SATELLITE AND GROUND-BASED OBSERVATION NETWORK

The CALIPSO (Cloud Aerosol Lidar and Infrared Pathfinder Satellite Observations) satellite was launched on April 28, 2006 as a joint NASA and CNES mission [9]. CALIPSO is part of a train of seven satellites (five of which are orbiting at the date of this article), known as the "A-Train," in sun-synchronous orbit at a mean altitude of approximately 690 km [21]. A typical 1-day ground track for CALIPSO is shown in Fig. 1; CALIPSO completes an orbit every 98.4 minutes (approximately 14.6 orbits per day), and repeats its track every 16 days. CALIPSO contains a LIDAR (Light Detection and Ranging) system, known as CALIOP (Cloud Aerosol Lidar with Orthogonal Polarization), with a primary mission of obtaining high-resolution vertical profiles of clouds and aerosols in the Earth's atmosphere [22]. The CALIOP laser produces simultaneous, co-aligned 20 ns pulses of 532 nm and 1064 nm light, pointed a small angle (0.3°) away from the geodetic nadir in the forward along-track direction, at a repetition rate of 20.16 Hz. The light enters a beam expander, following which the divergence of each laser beam wavelength is approximately 100 μrad, producing a spot of approximately 70 m RMS diameter on the ground. The pulse energy is monitored onboard the satellite, and averages approximately 110 mJ, at each one of the two wavelengths, per pulse.

During 2006-2009, the CALIPSO beam was observed at several locations in western North America using a portable ground station consisting of seven simultaneously shuttered Panasonic FZ50 digital cameras, connected as shown in Fig. 2, and two calibrated photodiodes co-located with the central camera. Observing locations were selected by means of the CALIPSO ground track, monitored by NASA. For each observation, a point along the ground track was selected the day prior to the night of the overpass by means of the following criteria: road accessibility, lack of obscuring trees and streetlights, avoidance of fenced-off private land, and local weather conditions. A desert environment is ideal; for this reason, several of the observations were performed in the southwestern U.S. Even thin cloud cover can create a difference in time-averaged optical density between the CALIPSO beam path and light from stellar standard sources on the image nearby; thus all observations were performed on clear nights.

## 4. CONCLUSIONS AND FUTURE

The analysis of the data obtained during the CALIPSO observation campaign is currently in progress, to be published shortly. We intend to demonstrate the viability of using a manmade source in space for absolute photometry, and future improvements that can be made using this technique.

Precision astrophysical and cosmological measurements are starting to be limited by the use of present-day calibration sources and techniques. In order to keep up with the progress in the quality of surveys in the upcoming decade and beyond, new technologies will be required to reduce the systematic uncertainties associated with calibration, which have started to form a "floor," slowing the pace of progress in multiple areas. In order to make large steps, beyond the slowing pace of progress obtained by improving stellar standards, towards improving these calibration-related uncertainties, new techniques, such as the use of precisely calibrated satellite-mounted light sources, are necessary.

## 5. ACKNOWLEDGEMENTS

The authors would like to acknowledge the critical help and advice from Dr. Christopher Stubbs at Harvard University, Dr. Susana Deustua at the Space Telescope Science Institute (STScI), Dr. Chris Pritchet at the University of Victoria, Dr. John McGraw at the University of New Mexico, and Dr. Yorke Brown. JA, KF, KF, KK, and MJ were supported by the Canada Foundation for Innovation / British Columbia Knowledge and Development Fund Grant #13075 and the Univ. of Victoria. JB was supported by the U.S. National Science Foundation grant AST-0507475.